\documentclass[prb,aps,showpacs,floatfix,floats,nobalancelastpage,twocolumn]{revtex4-1}
\usepackage{graphicx}
\usepackage{graphics}
\usepackage{latexsym,amsmath,amssymb,bm,euscript, bbm}
\usepackage{color}
\usepackage{dcolumn}
\usepackage{bm}
\usepackage{epstopdf}
\usepackage{times}
\usepackage{multirow}
\usepackage{wasysym}
\usepackage{subfigure}
\def\etal{~\textit{et~al.}}
\def\ra{\rangle}
\def\la{\langle}

\begin{document}

\title{Correlation effects in double-Weyl semimetals}
\author{Hsin-Hua Lai}
\affiliation{National High Magnetic Field Laboratory, Florida State University, Tallahassee, Florida 32310, USA}
\date{\today}

\begin{abstract}
We study the long-range Coulomb interaction effects on the double-Weyl fermion system which is possibly realized in the three dimensional semimetal HgCr$_2$Se$_4$ in the ferromagnetic phase. Within the one-loop renormalization group analysis, we find that there exists a stable fixed point at which the Coulomb interaction is screened anisotropically. At the stable fixed point, the renormalized Coulomb interaction induces nontrivial \textit{logarithmic} corrections to the physical quantities such as specific heat, compressibility, the electrical conductivity, and the diamagnetic susceptibility that are obtained utilizing the renormalization group equations.. 
\end{abstract}
\maketitle

\section{Introduction}
There has been recently much interest in semimetals, which support gapless quasiparticle excitations only in the vicinity of isolated band touching points in the Brillouin zone (BZ). When the Fermi energy is pinned to the band touching points, these semimetals can possess universal power-law behaviors for thermodynamic and transport quantities as a function of temperature or external frequency. There are many well known experimental examples of the semimetals which possess linearly dispersing massless Dirac quasiparticles in both two dimensions (2D) and three dimensions (3D), e.g. Monolayer graphene \cite{Novoselov_science,Novoselov_nature,PKim_Nature}  in 2D and Bi$_{1-x}$Sb$_x$ \cite{BLenoir1996,GAmit2007,graphenereview}, Pb$_{1-x}$Sn$_x$Te \cite{RDornhaus,PGoswami2011},  and Cd$_3$As$_2$ \cite{BSergey2014}, Na$_3$Bi \cite{Liu2014} in 3D. It is also possible to realize parabolic semimetals which possess parabolic dispersions at band touching, e.g. Berner-Stacked bilayer graphene \cite{Novoselov2006} in 2D, and HgTe \cite{RDornhaus}, gray tin \cite{SGroves}, and the normal state at high temperature for some 227 irradiates such as Pr$_2$Ir$_2$O$_7$ \cite{SNakatsuji2006,Machida2010, Abrikosov1974,Moon2013,IHerbut2014,HHLai2014} in 3D.

In the presence of strong spin-orbit interactions in three dimensions, the unusual semimetallic phase called the topological Weyl semimetals may exist and have been confirmed in TaAs recently \cite{Weyl-1st, Weyl-2nd, Weyl-3rd, Weyl-4th}. The Weyl semimetals are also predicted to exist in pyrochlore iridates \cite{PHosur2010,Wan2011, Witczak-Krempa2012}, cold atom systems \cite{KSun2011,JHJiang2012}, and multilayer topological insulator systems \cite{ABurkov2011,Halasz2012}. In close proximity of the gapless points, the effective Hamiltonian is described by a two-component wave-function termed the Weyl fermion and the gap closing point is the Weyl node. The Weyl nodes are protected from opening a gap against infinitesimal translations of the Hamiltonian; these points act as monopoles (vortices) of 3D Berry curvature, as any closed 2D surface surrounding one of them exhibits a finite Chern flux, and a Weyl node can only be gapped by annihilation with an anti-Weyl node of opposite monopole charge. 

In addition to the usual Weyl semimetals, recently Ref.~\onlinecite{CFeng2012} proposed the possible presence of new 3D topological semimetals in materials with point group symmetries termed as double-Weyl semimetals. The new double-Weyl semimetals possess Weyl nodes with \textit{quadratic} dispersions in two directions, e.g. $\hat{x}-\hat{y}$ plane. The double-Weyl nodes are protected by $C_4$ or $C_6$ rotation symmetry and are suggested to be realized in the 3D semimetal HgCr$_2$Se$_4$ in the ferromagnetic phase, which possess a pair of double-Weyl nodes along $\Gamma Z$ direction \cite{CFeng2012,XGang2011}. The first-principle calculations in the material HgCr$_2$Se$_4$ \cite{XGang2011} also suggested the existence of double-Weyl nodes, which is qualitatively in agreement with the recent transport experiments in HgCr$_2$Se$_4$ \cite{Guan2015} that confirm the half-metallic property of the HgCr$_2$Se$_4$. The (anti-)double-Weyl node possesses a monopole charge of (-2)+2 and the double-Weyl semimetal shows double Fermi-arcs on the surface BZ \cite{CFeng2012,XGang2011}. This new semimetallic phase with an in-plane quadratic dispersion can serve as a new platform for studying the (long-range) Coulomb interaction effects on the double-Weyl fermion. The low-energy physics of the double-Weyl fermion can possibly serve as a new source term contributing to the physical properties in HgCr$_2$Se$_4$ for chemical potential sitting near the Weyl nodes, such as a $T^2$ term to the specific heat $C$ that was not considered previously \cite{WDWang2013}.

In this work, we consider a single double-Weyl fermion coupled to the long-range Coulomb interaction and study the effects within the one-loop renormalization group (RG) analysis in the Wilsonian momentum shell scheme \cite{ShankarRG_RMP}. Due to the anisotropic dispersions of the double-Weyl node, the density of states (DOS) is linearly proportional to the energy, $D(\epsilon) \propto \epsilon$, sharply different to that in the usual Weyl fermion with $D(\epsilon) \propto \epsilon^2$. Due to the anisotropic dispersions, the scalings for the three spatial coordinates can be different. In the noninteracting limit, for scaling transformation invariance of the action we find the dynamical scaling exponent $z=2$, the scaling exponents of the spatial coordinates $\vec{x}$ and $\vec{y}$ are $1$, i.e. the scaling dimension $[\vec{x}]=[\vec{y}]=-1$, while that of $\vec{z}$ is two, $[\vec{z}] = - 2$. The result of such nontrivial scaling transformations in spatial dimensions result in the Coulomb interaction $e^2$ with engineering scaling dimension $[e^2] = z-1 = 1$ and anisotropy parameter $\eta$, which dictates the anisotropy of the system, with enginnering scaling dimension $[\eta] = -2$. 

After coarse-graining within RG analysis, we find that in the low-energy limit, the system becomes highly anisotropic and $\eta$ becomes infinitesimal. The Coulomb potential receives strong renormalization along the linear dispersion axis and the Coulomb interaction strength $e^2$ also becomes infinitesimal due to the strongly irrelevant anisotropy variable $\eta$. The composite variable similarly equal to the ratio of Coulomb interaction strength and the square root of the anisotropy parameter, $\sim e^2/\sqrt{\eta}$, approaches a \textit{fixed} value in the low-energy limit, which defines the \textit{stable} fixed point in RG. At the stable fixed point,  the Coulomb potential is renormalized anisotropically, which is consistent with the random phase approximation (RPA) calculation detailed in the supplemetal material. Furthermore, we find that the square of the Coulomb interaction,$\sim e^4$, under coarse-graining process decreases in a logarithmic manner, which leads to logarithmic suppressions to several physical quantities such as specific heat $C$ compressibility $\kappa$ and the so called finite frequency (dynamic) conductivity $\sigma_{\mu\mu=x,y,z}(\omega)$, and the dc conductivity $\sigma^{dc}_{\mu \mu}(T)$. Furthermore, we find unexpectedly that the diamagnetic susceptibility $\chi_D$ gets {\it enhanced} in a logarithmic manner due to the long-range Coulomb interaction.

The paper is organized as follows. In Sec.~\ref{Sec:model} we introduce the model Hamiltonian followed by weak-coupling RG analysis. In Sec.~\ref{Sec:logarithmic_corrections} we utilize the RG equations near the fixed point to obtain logarithmic corrections to various physical quatities. In Sec.~\ref{Sec:discussions} we conclude with some discussions.

\section{Double Weyl Semimetal in the long-range Coulomb interaction}\label{Sec:model}
We consider a minimal model of a single double-Weyl fermion coupled to the long-range Coulomb interaction. The action in the Euclidean path integral formalism is 
\begin{eqnarray}\label{Eq:action}
\nonumber S_L = && \int d \tau d^3 \vec{x}  \bigg{\{} \psi^\dagger \left[ \partial_\tau - i e \phi + \vec{d}(-i \nabla) \cdot \vec{\sigma} \right] \psi + \\
&& \hspace{1cm} + \frac{1}{2\sqrt{\eta}} \left[ (\partial_x \phi)^2 + ( \partial_y \phi)^2 \right] + \frac{\sqrt{\eta}}{2}(\partial_z \phi )^2 \bigg{\}},~~\label{Eq:action}
\end{eqnarray}
with $\vec{d} \equiv \left\{ - m^{-1} \left( \partial_x^2 - \partial_y^2\right),~ - m^{-1} (2 \partial_x \partial_y),~ -iv_z\partial_z\right\}$, where $\psi$ and $\phi$ represent the electron annihilation field and the boson annihilation field. The integration of $\phi$ gives the usual instantaneous long-range Coulomb interaction. The variable $m$ along the $\vec{x}-\vec{y}$ plane represents the effective mass and the variable $v_z$  along $\vec{z}$ direction is the velocity. The anisotropy variable $\eta$ is introduced due to the anisotropic dispersions. 

We choose the scaling transformations for the fields and the variables as $\tau = b^{z}\tau_R, ~x = b~x_R,~y= b~y_R,~z = b^{z_1}~z_R$, $v_z = Z_{v_z}^{-1} v_{z,R},~m^{-1} = Z_{m^{-1}}^{-1} m^{-1}_R,~\eta = Z_\eta^{-1} \eta_R$, $e = Z_e^{-1/2} e_R$, $\psi = Z_\psi^{-1/2} \psi_R,~\phi = Z_\phi^{-1/2} \phi_R$, where the parameter $b=e^{\ell} $ represents a length scale slightly greater than one with a logarithmic length scale $\ell \ll 1$ and the subscript $R$ labels renormalized variables during coarse-graining. 

\begin{figure}[t] 
   \centering
   \includegraphics[width=2.5in]{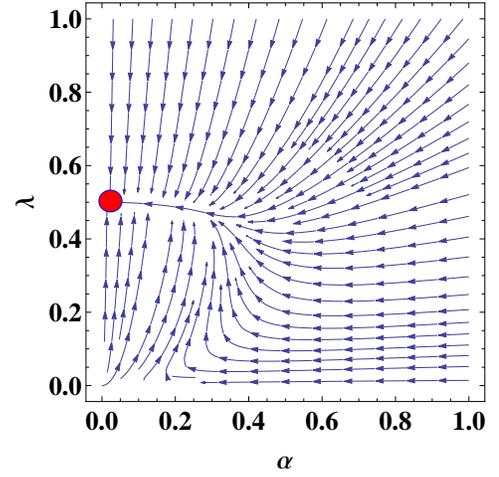} 
   \caption{RG stream plot for the model action, Eq.~(\ref{Eq:action}). The red dot represent the fixed point $(\alpha^*,\lambda^*) = (0,1/2)$, where the Coulomb interaction receives anisotropic screening. The RG streams mostly flow to the stable fixed point along the $\lambda =1/2$ path.}
   \label{Fig:DWS_RG}
\end{figure}

For analytically extracting the RG equations, we adopt the RG scheme in Ref.~\onlinecite{BJYang2014} and perform integration within a momentum shell in the $q_\perp \equiv \sqrt{q_x^2 +q_y^2}$ direction, i.e. $q_\perp \in [\Lambda e^{-\ell}, \Lambda ]$, and no restriction along $q_z$ direction $(|q_z| \in [0, \Lambda_z \rightarrow \infty])$. For clarity in the presentation of RG results, we introduce two variables $ \alpha \equiv \frac{me^2}{12\pi^2 \Lambda}$, and $\lambda \equiv \frac{m^2 v_z e^2}{48\pi^2 \sqrt{\eta}\Lambda^2}$. We obtain the one-loop RG equations, detailed in App.~\ref{APP:RG}, 
\begin{eqnarray}
&& \frac{d \ln v_z}{d\ell} = z - z_1 +\frac{3}{4}\frac{\alpha^2}{\lambda},\\
&& \frac{d \ln m^{-1}}{d\ell} = z -2 + \frac{3c}{4}\frac{\alpha^2}{\lambda},\\
&& \frac{d\alpha}{d\ell} = \alpha \left[ 1 - 2\lambda  - \frac{3c+2}{4}\frac{\alpha^2}{\lambda}\right],\\
&& \frac{d \lambda}{d\ell} = 2\lambda  \left[ 1 - 2\lambda  - \frac{6c-3}{8}\frac{\alpha^2}{\lambda}\right].
\end{eqnarray}
where $c = \ln (3+2\sqrt{2})/2 \simeq 0.881$. If we hold $v_z$ and $m^{-1}$ fixed, we get
\begin{eqnarray}
&& z= 2 - \frac{3c}{4} \frac{\alpha^2}{\lambda},\\
&& z_1 = z + \frac{3}{4}\frac{\alpha^2}{\lambda} = 2 + \frac{3(1-c)}{4}\frac{\alpha^2}{\lambda}.
\end{eqnarray}
We can see that fixed points are located at $( \alpha,~\lambda) = (0,~0)$, and $(0,~1/2).$ Linearizing around these two fixed points, we find that the fixed point $(0,~0)$ is the {\it unstable} Gaussian fixed point, and $(0,~1/2)$ is the {\it stable} fixed point controlled by the parameters $\alpha$ and $\lambda$. The RG streamplot is shown in Fig.~\ref{Fig:DWS_RG}. The Coulomb interaction decreases to the stable fixed point mostly along the path of $\lambda = 1/2$. Along this path, the square of the Coulomb interaction ($e^4$) decreases to zero in a \textit{logarithmic} manner, which is reflected as the nonmonotonic temperature or frequency dependences of physical quantities such as the specific heat $C$, compressibility $\kappa$, the finite frequency (dynamic) conductivity $\sigma_{\mu\mu = x,y,z}(\omega)$, the dc conductivity $\sigma^{dc}_{\mu \mu}$, and the diamagnetic susceptibility $\chi_D$, which we will show below.

At the fixed point $(\alpha_s, \lambda_s) = (0, 1/2)$, the dynamical exponent $z = 2 = z_1$. If we focus on the renormalized boson propagator (which gives the Coulomb screening effects), at the fixed point it is (below we will suppress the irrelevant dimensionful variables)
\begin{eqnarray}
\nonumber q_\perp^2 + q_z^2 - \Pi(q) & \sim& q_\perp^2 ( 1 + \alpha_s^2/\lambda_s  \ell) + q_z^2 (1 + 4 \lambda_s \ell) \\
&=& q_\perp^2 + q_z^2 (1 + 2\ell),
\end{eqnarray}
which shows that there is only a correction along the $q_z$ direction. If we consider the relative scalings between $q_\perp$ and $q_z$, we can obtain
\begin{eqnarray}
q_\perp^2 + q_z^2 - \Pi(q) \sim q_\perp^2 + q_z^{2}|q_z|^{-\frac{2}{z_1}}=q_\perp^2 + |q_z|,
\end{eqnarray}
where $z_1 =2$ at the stable fixed point. The renormalized Coulomb interaction at the fixed point becomes $V_{c}(\vec{q}) \sim 1/(q_\perp^2 + |q_z|)$. Fourier transforming $V_c(\vec{q})$ back to the real space, we find that the renormalized Coulomb potential falls off anisotropically, $V_{c}(r_\perp, |z|=0) \sim r_\perp^{-2}$, and $V_c(r_\perp = 0, |z|) \sim |z|^{-1}$, where $r_\perp \equiv \sqrt{x^2+y^2}$. In the App.~\ref{APP:RPA}, we perform RPA analysis of the screened Coulomb interaction and we find that the scaling analysis above is consistent with the RPA analysis.

\section{Logarithmic corrections to the scaling behavior of physical quantities}\label{Sec:logarithmic_corrections}
According to the strong-coupling analysis in App.~\ref{APP:Large_N}, we find that the renormalized Coulomb interaction causes an infrared logarithmic divergence that can lead to the logarithmic corrections to physical quantities. Instead of calculating higher-order corrections in the perturbation theory, we follow Ref.~\onlinecite{Sheehy2007} to utilize the RG equations near the stable fixed point and the scaling arguments to obtain the scaling behaviors of the physical quantities. 

Focusing on the path of $\lambda =1/2$ near the stable fixed point, we know that the mass inverse $m^{-1}$ and velocity $v_z$ receive nontrivial renormalization as $m^{-1} ( 1 + \frac{3c}{2}\alpha^2 \ln b)$ and $v_z ( 1 + \frac{3}{2} \alpha^2 \ln b)$, where $\ln b = \ell$. From the scaling invariance of action, we know $Z_\psi = b^{z_1 +2}$, $Z_{m^{-1}} = b^{z-2} \left( 1 + \frac{3c}{2}\alpha^2 \ln b\right)$, and $Z_{v_z} = b^{z-z_1} \left( 1 + \frac{3}{2} \alpha^2 \ln b \right)$. For a finite potential term, we can consider to add a term $ - \mu \int d^3x \int d\tau \psi^\dagger\psi$ to the action and we can obtain the transformation $\mu = b^{-z} \mu_R$. We can also consider the free energy $F$, which can be a general function of $T$, $\alpha$, $\mu$, magnetic field $B$, etc., that transforms as $F = Z_F^{-1} F_R$. Considering the exponent of a partition function $
\int d^3 x \int d\tau F$, we know that the exponent should be dimensionless, which requires $F = b^{-(2+z_1)}b^{-z} F_R$. In addition, the temperature transforms under coarse-graining as $T = b^{-z} T_R$.

The relevant RG equations near the stable fixed point along the path $\lambda =1/2$ are
\begin{eqnarray}
&& \frac{d\alpha}{d \ln b} = -\frac{3c+2}{2} \alpha^3(b),\\
&& \frac{dT}{d\ln b} = T(b) \left[ 2 - \frac{3c}{2}\alpha^2(b) \right].
\end{eqnarray}
Solving the RG equations, we get
\begin{eqnarray}
\alpha^2(b) & = &\alpha^2 \left( 1 + (3c+2)\alpha^2 \ln b\right)^{-1},\\
\nonumber T(b) &= &T b^2 \left( 1 + (3c+2)\alpha^2 \ln b\right)^{-\frac{3c}{2(3c+2)}} \\ 
&\simeq & T b^2 \left( 1 + \frac{3c}{2} \alpha^2 \ln b \right)^{-1},
\end{eqnarray}
where $\alpha$ and $T$ represent the bare Coulomb interaction $\alpha(0)$ and bare temperature $T(0)$. Choosing the temperature cut-off $T(b^*) =T_0 = m^{-1} \Lambda^2$, we get
\begin{eqnarray}\label{Eq:T-cutoff}
b^* \simeq  \left[ \frac{T_0}{T} \left( 1 + \frac{3c}{4}\alpha^2 \ln \frac{T_0}{T} \right) \right]^{\frac{1}{2}}
\end{eqnarray}
The renormalization of specific heat under RG can be obtained as $
C = - T \frac{\partial^2 F}{\partial T^2} = b^{-(2+z_1)} C_R.$ Choosing the cut-off $b$ to be $b^*$ and using the high temperature result of specific heat, $C_R \sim T_0^2$ which is originated from $D(\epsilon) \sim \epsilon$ in the noninteracting double-Weyl semimetals, we get
 \begin{eqnarray}\label{Eq:specific_heat}
 \nonumber C\sim \frac{T^2}{\left( 1 + \frac{3}{4}\alpha^2 \ln \frac{T_0}{T}\right) \left( 1 + \frac{3c}{4}\alpha^2 \ln \frac{T_0}{T}\right)}\sim \frac{T^2}{\left(1+ \frac{3}{4} \alpha^2 \ln \frac{T_0}{T}\right)^2},\\
 \end{eqnarray}
 where we crudely approximate $c\sim 1$ in the last line. The compressibility can be obtained as $\kappa  \equiv \frac{\partial^2 F}{\partial \mu^2} = b^{z} b^{-(2+z_1)} \kappa_R$. If we substitute $b^*$ for $b$ and use the noninteracting result $\kappa_R \sim T_0$, we get
 \begin{eqnarray}
 \nonumber \kappa \sim \frac{T}{\left( 1 + \frac{3}{4}\alpha^2 \ln \frac{T_0}{T}\right) \left( 1 + \frac{3c}{4}\alpha^2 \ln \frac{T_0}{T}\right)} \sim \frac{T}{\left(1+ \frac{3}{4} \alpha^2 \ln \frac{T_0}{T}\right)^2}.\\
 \end{eqnarray}
 
Let's shift our focus on the scaling behavior of the finite frequency conductivity and dc conductivity. In the presence of magnetic vector potential, the kinetic terms are modified as $-i\partial_j \rightarrow -i \partial_j + eA_j$, with $j = x,~y,~z$. Due to the minimal coupling, we require that the composite variables $eA_j$ rescale the same as that of $\partial_j$, which leads to $A_\perp = b ^{-1}\left[ 1 + \frac{3c}{2}\alpha^2 \ln b\right]^{-1/2}\left[1+ 2 \alpha^2 \ln b\right]^{-1/4}A_{\perp,R}$, with $A_\perp = A_{x/y}$, and $A_z = b^{-2} \left[ 1 + \frac{3c}{2}\alpha^2 \ln b\right]^{1/2}\times$ $\times\left[ 1 + \frac{3}{2}\alpha^2 \ln b \right]^{-1} \left[ 1 + 2 \alpha^2 \ln b \right]^{-1/4}A_{z,R}$, where we explicitly use $\lambda = 1/2$ near the fixed point and the scaling of electric charge $Z_e$ obtained in the RG equations derivation in App.~\ref{APP:RG}.

In order to obtain the scaling relations for the conductivity, we can rely on the Kubo formula. According to the Kubo formula, the current-current correlation function $\Pi_{\mu \mu}(i\omega_n) = \int d\tau e^{i\omega_n \tau} \int d^3 q \left\la T_\tau \left[ j_\mu (\vec{q},\tau) j_\mu (\vec{q}, 0) \right] \right\ra$ can be related to the dynamic conductivity as $\sigma_{\mu \mu}(\omega) = - Im \Pi_{\mu \mu}(\omega)/\omega$, where $i\omega_n \rightarrow \omega + i 0^+$. $j_\perp (\vec{q},\tau)$ and $j_z(\vec{q},\tau)$ can be obtained from the Fourier transform of $j_\perp (\vec{x},\tau) \simeq e m^{-1} \psi^\dagger (\vec{x},\tau) (-i\nabla) \cdot \vec{\sigma}_{\perp} \psi(\vec{x},\tau)$ and $j_z(\vec{x},\tau) = e v_z \psi^\dagger(\vec{x},\tau) \sigma^z \psi (\vec{x},\tau).$
We first obtain that $j_\perp (\vec{q},\tau) = b^{-1} \left[1+\frac{3c}{2}\alpha^2\ln b\right]^{1/2}\left[1+2\alpha^2\ln b\right]^{1/4} j_{\perp, R}$ and $j_z (\vec{q},\tau) = \left[ 1 + \frac{3c}{2}\alpha^2 \ln b\right]^{1/2} \left[ 1 + 2\alpha^2 \ln b\right]^{1/4} j_{z,R}$, which lead to $\sigma_{\perp\perp} = b^{-2}\left[1+\frac{3}{2}\alpha^2\ln b\right]^{-1}\left[ 1 + 2\alpha^2 \ln b \right]^{1/2} \sigma_{\perp \perp,R}$, and $\sigma_{zz} = \left[ 1 + \frac{3c}{2}\alpha^2 \ln b\right]^{-1}\left[ 1 + 2\alpha^2 \ln b\right]^{1/2} \sigma_{zz,R}.$
For $\omega > T$ (with $T\rightarrow 0, \omega\rightarrow 0$), we introduce the cut-off frequency $\omega_0 = \omega(b^*)$ with 
\begin{eqnarray}
b^* \simeq  \left[ \frac{\omega_0}{\omega} \left( 1 + \frac{3c}{4}\alpha^2 \ln \frac{\omega_0}{\omega} \right) \right]^{\frac{1}{2}},
\end{eqnarray}
which is due to the fact that the $\omega(b)$ scales the same as the temperature $T(b)$, i.e. $ \mathcal{O} = b^{-z} \mathcal{O}_R$, where $\mathcal{O} = T,~\omega$. With the cut-off frequency, we obtain
\begin{eqnarray}
\nonumber  \sigma_{\perp\perp}(\omega) &\sim& \frac{\left[1+\alpha^2\ln\left(\frac{\omega_0}{\omega}\right)\right]^{\frac{1}{2}}}{\left[1+\frac{3c}{4}\alpha^2\ln\left(\frac{\omega_0}{\omega}\right)\right]\left[1+\frac{3}{4}\alpha^2\ln \left(\frac{\omega_0}{\omega}\right)\right]}e^2\omega \label{Eq:dynamic_C1}\\
&\simeq& \left[1 - \frac{3c + 1}{4}\alpha^2 \ln\left( \frac{\omega_0}{\omega}\right)\right]e^2\omega,\\ 
\nonumber \sigma_{zz}(\omega) &\sim& \frac{\left[1+\alpha^2\ln \left(\frac{\omega_0}{\omega}\right)\right]^{\frac{1}{2}}}{1+\frac{3}{4}\alpha^2\ln\left(\frac{\omega_0}{\omega}\right)}e^2 \simeq \left[1-\frac{1}{4}\alpha^2\ln\left(\frac{\omega_0}{\omega}\right)\right]e^2.\\ \label{Eq:dynamic_C2}
\end{eqnarray}
where we approximate $\sigma_{\perp \perp,R}(\omega)$ and $\sigma_{z,R}(\omega)$ to be the noninteracting results of the dynamic conductivity at finite frequency obtained in App.~\ref{APP:dynamic_conductivity}. 

The dynamic conductivity calculations at $\mu\ll T$ in the noninteracting limit in App.~\ref{APP:dynamic_conductivity}l also show the existence of the Drude peak and the linear-$T$ dependent $xx/yy$-component dc conductivity, $\sigma^{dc}_{xx/yy}\equiv \sigma^{dc}_{\perp \perp}$, and $T$-independent $zz$-component dc conductivity, $\sigma^{dc}_{zz}$. Following similar discussions above with high temperature cut-off, Eq.~(\ref{Eq:T-cutoff}), the dc conductivity also receive nonmonotonic temperature suppression, which are similar to Eqs~(\ref{Eq:dynamic_C1})-(\ref{Eq:dynamic_C2}) with $\omega\rightarrow T$,
\begin{eqnarray}
&& \sigma^{dc}_{\perp \perp}(T) \sim \left[1 - \frac{3c + 1}{4}\alpha^2 \ln\left( \frac{T_0}{T}\right)\right] e^2 T,\\ 
&& \sigma^{dc}_{zz}(T) \sim \left[1-\frac{1}{4}\alpha^2\ln\left(\frac{T_0}{T}\right)\right]e^2.
\end{eqnarray}

Last but not least, we focus on the temperature dependence of the diamagnetic susceptibility. The diamagnetic susceptibility can be obtained from taking second derivative of the free energy, $\chi_D = - \partial^2 F/\partial B^2$. Since $\vec{B} = \nabla \times \vec{A}$, the renormalization of the magnetic field under coarse-graining can be obtained straightforward using the renormalization of $A_j$ illustrated above. We find that the diamagnetic susceptibility scales differently for in-plane magnetic field $\vec{B} = B \vec{r}_\perp$ and for perpendicular magnetic field $\vec{B} = B\hat{z}$. For $\vec{B} = B \vec{r}_\perp$, the diamagnetic susceptibility renormalizes as $ \chi_D^{\perp} = \left[ 1 + \frac{3c}{2}\alpha^2 \ln b \right] \left[ 1 + \frac{3}{2}\alpha^2 \ln b\right]  \left[ 1 + 2 \alpha^2 \ln b\right]^{1/2}\chi_{D,R}^{\perp}.$  For $\vec{B} = B \hat{z}$, the diamagnetic susceptibility renormalizes as $\chi_D^z =b^{-2} \left[ 1 + \frac{3c}{2}\alpha^2 \ln b\right]^3\left[1 + \frac{3}{2}\alpha^2 \ln b \right]^{-1} \left[ 1 + 2 \alpha^2 \ln b\right]^{1/2}\chi_{D,R}^z$.

We use Eq.~(\ref{Eq:T-cutoff}) and the noninteracting results of $\chi_{D,R}$ which we derive using the Fukuyama formula for the orbital diamagnetic susceptibility \cite{Fukuyama1971} in App.~\ref{APP:diamagnetic_sus}. We obtain
\begin{eqnarray}
\chi_D^\perp&\sim& \left[ 1 + \frac{3c+5}{4}\alpha^2 \ln \left(\frac{T_0}{T}\right)\right]e^2 v_z,\\
 \chi_D^{z} &\sim& \left[ 1 + \frac{6c-2}{4}\alpha^2\ln \left(\frac{T_0}{T}\right)\right] \frac{e^2T}{mv_z}.
\end{eqnarray}
We expect that the temperature dependence of the diamagnetic susceptibility for a magnetic field in general direction at low temperature to be  
\begin{eqnarray}
\chi_D \sim \left[ 1 + \alpha^2 \ln \left( \frac{T_0}{T}\right)\right] \left( \sin^2\theta~\chi_0 + \cos^2\theta~T\right),
\end{eqnarray}
where $\theta$ is the angle between the magnetic field and $\vec{z}$-axis, i.e. $\vec{B}\cdot\vec{z} = B \cos\theta$, and $\chi_0$ is a constant independent of temperature and the diamagnetic susceptibility gets \textit{enhanced}.

\section{Discussions}\label{Sec:discussions}
We study the long-range Coulomb interaction effects on the double-Weyl semimetals. Within one-loop renormalization group analysis we find that the composite variable defined as the ratio of the Coulomb interaction strength and the square root of the anisotropy parameter, $\sim e^2/\sqrt{\eta}$, is fixed to be finite at long-wavelength, which defines the fixed point. Focusing near the fixed point, we utilize RG equations to obtain nonmonotonic temperature or frequency dependences of various physical quantities.

Though the long-range Coulomb interaction induces logarithmic corrections to several physical quatities in experiment, the fundamental Berry curvature structure around the double Weyl-point remains unaltered, similar to the situations in single-Weyl point and the Dirac points of graphen \cite{Krempa2014}. In the presence of Coulomb interaction, the renormalized low-energy description near a double-Weyl point is similarly $H_f(\vec{ k}) \sim \left[ 1+\alpha^2 \ln\left(\Lambda/|k|\right)\right] \vec{d}(\vec{k})\cdot \vec{\sigma}$. The Berry curvature $\nabla \times \vec{A}$ is independent of the overall real renormalization factor since it measures the ``complex phase'' of the Hamiltonian eigenstates as they are parallel transported in the BZ. The Chern flux through a small sphere enclosing a double-Weyl point remains unaltered and so do the associated topological quantities.

Despite the similarities between the present work and Ref.~\onlinecite{BJYang2014}, the conclusions are in sharp difference. The RG fixed point in Ref.~\onlinecite{BJYang2014} is defined, in our convention, as the composite parameter $\sim e^2\sqrt{\eta}$ flowing to a fixed value with $e^2 \rightarrow 0$ and $\sqrt{\eta}\rightarrow \infty$. In stark contrast to our results, the $e^2$ in Ref.~\onlinecite{BJYang2014} vanishes \textit{exponentially} under RG flow, which leads to the fact that the physical properties, such as $C$, $\kappa$, and etc., are the same to \textit{noninteracting} ones. 

In the end, we briefly discuss the effects of the short-range interactions and disorders. The Lagrangian density of a short-range interaction can be written similarly as $g_j (\psi^\dagger \Gamma_j \psi)^2$, where $\Gamma_j = \mathbbm{1}_2, \sigma_i$. A short-range coupling at tree-level is stronly irrelevant and scales as $g_j =b^{-2} g_{j,R}$. A $e^4$ term may be generated under RG that drives the short-range couplings to strong coupling, similar to the situations in the parabolic semimetals \cite{IHerbut2014, HHLai2014}. However, due to the fact that the $e^2$ vanishes near the fixed point, the short-range couplings remain irrelevant and negligible. The effects of the disorders are more intriguing and detrimental. The Lagrangian density of a disorder can be written as $V_j \psi^\dagger M_j \psi$, with $M_j = \mathbbm{1}_2, \sigma_j$. If we choose Gaussian white noise distribution for the disorder according to $\la\la V_i (\vec{x}) V_j(\vec{x}') \ra\ra = \Delta_{ij} \delta^{(3)}(\vec{x}-\vec{x}')$, we perform the average over disorder by employing replica method \cite{PGoswami2011, HHLai2014}. The effective disorder terms mimic the four-fermi interactions but nonlocal in imaginary time, i.e. $\bar{S}_D \sim \int d^3\vec{x} d\tau d\tau' \Delta_j \psi^\dagger_a (\vec{x},\tau) M_j \psi_a (\vec{x},\tau) \psi^\dagger_b (\vec{x},\tau') M_j \psi_b(\vec{x},\tau')$, where $a,b$ are replica indices. The disorder average vertices are \textit{marginal}, $\Delta_j = \Delta_{j,R}$, at the tree-level RG analysis. Hence, a more thorough treatment including one-loop corrections is needed, which we leave for the future studies.\\

 \textit{Note added}--During the journal review process, we found a preprint \cite{Yao_DWS} working on similar topic. Ref.~\onlinecite{Yao_DWS} also finds a new fixed point within one-loop weak-coupling RG analysis in the presence of the long-range Coulomb interaction, where the Coulomb interaction gets screened anisotropically and specific heat, $C$, receives a logarithmic correction, similar to the conclusion of the present paper. However, the qualitative differences between the results here and those in Ref.~\onlinecite{Yao_DWS} originate from the different way of introducing anisotropy, the second term in Eq.~(\ref{Eq:action}) involving the bosonic field $\phi$ and the anisotropy variable $\eta$. Due to the subtle difference of characterizing the anisotropy, the fixed points are different. Unlike the RG fixed point in this paper defined as $\lambda \sim e^2/\sqrt{\eta} \rightarrow constant$, the RG fixed point in Ref.~\onlinecite{Yao_DWS} is defined as $\sim e^2/\eta \rightarrow constant$, which leads them to the result that the logarithmic correction to the specific heat is $\delta C \sim -T^2\alpha \ln (T_0/T)$, in constrast to our result in Eq.~(\ref{Eq:specific_heat}), $\delta C\sim - T^2 \alpha^2\ln (T_0/T)$. Very recently, the double-Weyl semimetal is also proposed to be realized in SrSi$_2$ \cite{DWS_proposal}.
  
\acknowledgments
H.-H. Lai thanks J. Murray, G. Chen, P. Goswami, and K. Yang. This work is supported by the National Science Foundation through grants No. DMR-1004545 and No. DMR-1442366.
\appendix

\section{One-loop RG corrections for double-Weyl Semimetal in clean limit}\label{APP:RG}
The action of a double-Weyl fermion coupled to the long-range Coulomb interaction is given in Eq.~(\ref{Eq:action}) in the main texts. We can define the fermion Green's function and the (boson) scalar Green's function as
\begin{eqnarray}
&& G_0(\vec{k},\omega)  =  \frac{i\omega+\vec{d}\cdot \vec{\sigma}}{\omega^2 + m^{-2} k_\perp^4+v_z^2 k_z^2},~~~\\
&& D_0(\vec{k},\omega)  = \frac{\sqrt{\eta}}{k_\perp^2 +\eta k_z^2},
\end{eqnarray}
where we define $\vec{k}_\perp = ( k_x, k_y)$ and $|\vec{k}_\perp| \equiv k_\perp \equiv \sqrt{k_x^2 +k_y^2}$.
\begin{figure}[t] 
   \centering
   \includegraphics[width=2.5in]{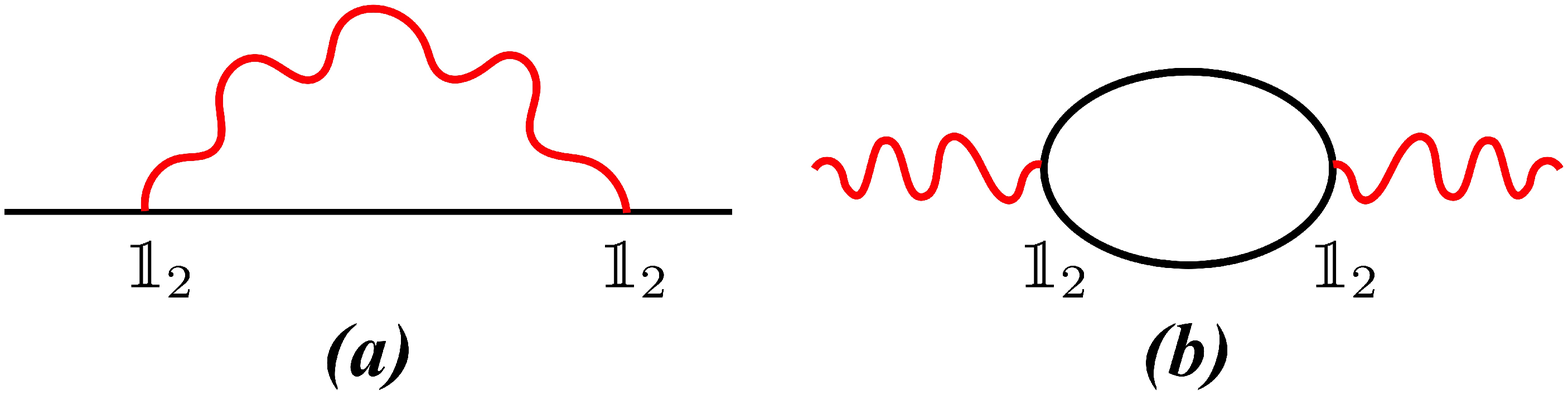} 
   \caption{Feynman diagrams for the self-energy corrections due to the long-range Coulomb interaction. The red curvy lines are boson propagators introduced for performing Hubbard-Stratonovich transformation of the four-fermion Coulomb interaction. The blue lines are fermion propagators. The boson-fermion vertex is $-ie \mathbbm{1}_2$.}
   \label{Fig:Coulomb_FD}
\end{figure}
The Fig.~\ref{Fig:Coulomb_FD}(a) illustrates the Coulomb interaction induced fermion self-energy $\Sigma^{ex}(\vec{k},\omega)$ 
\begin{eqnarray}
\nonumber && \Sigma^{ex}(\vec{k},\omega) =  - e^2 \int_\Omega \int'_q  G_0 (q, \Omega)D_0(\vec{k}-\vec{q},\omega-\Omega)\\
&& =  \frac{ e^2}{2} \int'_q \frac{\vec{d}(\vec{q}) \cdot \vec{\sigma}}{\sqrt{m^{-2} q_\perp^4 + v_z^2 q_z^2}} \frac{\sqrt{\eta}}{|\vec{k}_\perp - \vec{q}_\perp |^2 + \eta (k_z - q_z)^2}.~~
\end{eqnarray}
where we introduce the abbreviations $\int_\Omega \equiv \int_{-\infty}^{\infty} d\Omega/(2\pi)$ and $\int'_q \equiv \int' d^3q/(2\pi)^3$, and the prime means the momentum integral within a momentum shell between $[ \Lambda e^{-\ell}, \Lambda]$, with $\ell \ll 1$. We adopt the RG scheme introduced by B.-J. Yang \etal~and introduce the large momentum cut-off $\Lambda$ along $\vec{q}_\perp$ direction, while there is no restriction for the integral along $q_z$. 
We take the calculation for the correction to $d_3\sigma^z$ for example. We consider $ Tr[\sigma^z \partial_{k_z}\Sigma^{ex}(\vec{k},0)]/Tr[\sigma^z \sigma^z]|_{\vec{k}\rightarrow 0}$, which gives the correction to $d_3(\vec{k})$,
\begin{eqnarray}
\nonumber  && \frac{Tr[\sigma^z \partial_{k_z}  \Sigma^{ex}(\vec{k},0)]}{Tr[\sigma^z \sigma^z]}\bigg{|}_{\vec{k}\rightarrow 0} \\
\nonumber&&= \eta^{3/2}v_z e^2 \int'_q \frac{q_z^2}{\sqrt{m^{-2}q_\perp^4+v_z^2 q_z^2} (q_\perp^2 + \eta q_z^2)^2} \\
\nonumber  && = \frac{\sqrt{\eta}e^2\ell}{4\pi^2} \int_{-\infty}^\infty dz \frac{z^2}{\sqrt{z^2 + \left( \frac{\sqrt{\eta} \Lambda}{m v_z}\right)^2} \left( z^2 + 1 \right)^2}  \\
\nonumber && = \frac{\sqrt{\eta}e^2\ell}{2\pi^2 m v_z}  \int_0^{\infty}  dz  \frac{1}{\sqrt{z^2 + A^2} \left( z^2 + 1 \right)^3}\\
&&\simeq \frac{\sqrt{\eta}e^2}{4\pi^2}\ell,
\end{eqnarray}
where we introduced $z=\sqrt{\eta}q_z/\Lambda$, and dimensionless $A \equiv \sqrt{\eta} \Lambda/(m v_z)$. If we solve the RG equations, we will see that the parameter $A$ is irrelevant and flows toward zero, and we show the leading contribution in the last line above.

For the correction to $d_2 \sigma^y$ for example. We consider $Tr[\sigma^y \partial_{k_x}\partial_{k_y}\Sigma^{ex}(\vec{k},0)]/Tr[\sigma^y \sigma^y]|_{\vec{k}\rightarrow 0}$, which will give the correction to $d_2(\vec{k})$, 
\begin{eqnarray}
\nonumber  &&\frac{Tr[\sigma^y \partial_{k_x} \partial_{k_y} \Sigma^{ex}(\vec{k},0)]}{Tr[\sigma^y \sigma^y]}\bigg{|}_{\vec{k}\rightarrow 0}\\
\nonumber && = \frac{8 \sqrt{\eta}e^2}{m} \int'_q \frac{q_x^2 q_y^2}{\sqrt{ m^{-2} q_\perp^4 + v_z^2 q_z^2} \left( q_\perp^2 + \eta q_z^2 \right)^3} \\
\nonumber &&= \frac{ \sqrt{\eta}e^2}{4\pi^2 m} \int_{\Lambda e^{-\ell}}^\Lambda q_\perp dq_\perp  \int_{-\infty}^\infty dq_z \frac{ q_\perp^4 }{\sqrt{m^{-2} q^4_\perp + v_z^2 q_z^2}\left( q_\perp^2 + \eta q_z^2 \right)^3}\\
\nonumber  &&= \frac{\sqrt{\eta}e^2}{2\pi^2 m v_z} \int_0^{\Lambda} dq_z \frac{\Lambda^6 \ell}{\sqrt{q_z^2 + \left( \frac{\sqrt{\eta} \Lambda^2}{m v_z}\right)^2} \left( q_z^2 + \Lambda^2 \right)^3}  \\
\nonumber && =\frac{\sqrt{\eta}e^2\ell}{2\pi^2 m v_z}  \int_0^{\sqrt{\eta}}  dz  \frac{1}{\sqrt{z^2 + A^2} \left( z^2 + 1 \right)^3}\\
&&=\frac{c \sqrt{\eta}e^2\ell}{2\pi^2 m v_z} ,
\end{eqnarray}
where $c = \ln (3+2\sqrt{2})/2\simeq 0.881$. During the calculation, we introduced the momentum cut-off for $q_z$ since if there is no restriction, we will get an artificial logarithm of $A$. It is also physically intuitive to introduce a momentum cut off for $q_z$. Since if we choose $\Lambda_\perp$ to be the largest momentum scale and perform the momentum shell integral along $q_\perp$, the largest momentum $\Lambda_z$ along $q_z$ should satisfy $v_z \Lambda_z \simeq m^{-1} \Lambda_\perp^2$, which are the energies along $q_\perp$ and $q_z$. Therefore, we can get an identity as $mv_z/(\Lambda_\perp^2/\Lambda_z) \simeq 1$. For $\Lambda_\perp \simeq \Lambda_z$, we can get $mv_z/\Lambda \simeq 1$ and thus $A = \sqrt{\eta} \Lambda/(mv_z) \simeq \sqrt{\eta}$. From rotation symmetry, we know the correction to $d_1(\vec{k})$ is the same to that of $d_2(\vec{k})$. Combining the corrections with the bare terms, we get
\begin{eqnarray}
\nonumber && \vec{d}_\perp(\vec{k})\cdot\vec{\sigma}_\perp \left[ 1 + \frac{ \sqrt{\eta} e^2 }{4\pi^2 v_z}\ell \right]+ d_3(\vec{k}) \sigma^z\left[1+\frac{\sqrt{\eta}e^2\ell}{4\pi^2v_z}\right],\\
\end{eqnarray}
where we define $\vec{\sigma}_\perp = ( \sigma_1, \sigma_2)$.  It may seem that there is some inconsistency in performing the RG calculations. The more valid way to perform the calculation should be stated as follows. Since we set $\Lambda_\perp$ to be the largest momentum scale and we perform momentum shell integration within $|q_\perp| \in [\Lambda_\perp e^{-\ell},\Lambda_\perp]$, we should consistently introduce the large momentum cutoff for the integrations of $q_z$ in the self-energy calculations. However, we note that these will simply complicate the coefficients of the corrections and the structure of the RG equations will remain \textit{the same}, i.e. the fixed point structure will remain \textit{the same}.

The Fig.~\ref{Fig:Coulomb_FD}(b) represents the boson self-energy $\Pi(\vec{k},\omega)$
\begin{eqnarray}
&& \Pi(\vec{k},\omega) = e^2 \int_\Omega \int'_q Tr \left[ G_0(\vec{q},\Omega) G_0(\vec{q}+ \vec{k},\Omega + \omega ) \right].
\end{eqnarray}
Since the boson propagator is frequency dependent, we can focus on static $\Pi (\vec{k}, \omega =0)$. After frequency integral, we get
\begin{eqnarray}
\nonumber \Pi(\vec{k},0) = -e^2 \int'_q \left[ \frac{1}{E_q + E_{k+q}}  - \frac{\vec{d}(\vec{q}) \cdot \vec{d}(\vec{q} + \vec{k})}{E_{q+k} E_q ( E_{q+k} + E_q) }\right],\\
\end{eqnarray}
where we define $E_q^2 \equiv m^{-2} q_\perp^4 + v_z^2 q_z^2$. After expansion to quadratic order in $\vec{k}$, the integrals give 
\begin{eqnarray}
\frac{e^2 \ell }{6 \pi^2 v_z }  \left( k_x^2 + k_y^2 \right) + \frac{m^2 v_ze^2\ell}{24\pi^2 \Lambda^2} k_z^2
\end{eqnarray}
Combining the corrections and the bare terms, we obtain
\begin{eqnarray}
\nonumber && \frac{1}{2\sqrt{\eta}}\left(k_x^2 + k_y^2 \right) \left[ 1 + \frac{\sqrt{\eta}e^2}{3\pi^2 v_z}\ell \right]  + \frac{\sqrt{\eta}}{2} k_z^2 \left[ 1 + \frac{m^2 v_z e^2}{12\pi^2 \sqrt{\eta} \Lambda^2 } \right].\\
\end{eqnarray}

The renormalized action after inclusion of the self-energy corrections due to the long-range Coulomb interaction is
\begin{widetext}
\begin{eqnarray}
\nonumber \tilde{S}_L = && \int d\tau d^3\vec{x}  \bigg{\{}  \psi^\dagger \bigg{[} \partial_\tau - i e \phi + \left( 1 + \frac{c \sqrt{\eta}e^2}{4\pi^2 v_z}\ell\right) \vec{d}_\perp \cdot \vec{\sigma}_\perp   + \left( 1 + \frac{\sqrt{\eta}e^2\ell}{4\pi^2v_z}\right)  d_3 \sigma_3 \bigg{]} \psi + \\
 && \hspace{1.2 cm}+ \frac{1}{2\sqrt{\eta}} \left( 1 + \frac{\sqrt{\eta}e^2}{3\pi^2 v_z}\ell\right) \bigg{[} (\partial_x \phi)^2 + ( \partial_y \phi)^2 \bigg{]} + \frac{\sqrt{\eta}}{2}\left(1 + \frac{m^2 v_z e^2\ell}{12\pi^2 \sqrt{\eta}\Lambda^2}\right) (\partial_z \phi )^2 \bigg{\}}.
\end{eqnarray}
\end{widetext}
We rescale the parameters as $\tau = \tau_R e^{z\ell}$, $x=x_Rb^{\ell}$, $y= y_R e^\ell$, $z = z_R e^{z_1\ell}$, $e=Z_e^{-1/2}e_R$,$\psi = Z_\psi^{-1/2} \psi_R,$ and $\phi = Z_\phi^{-1/2} \phi_R$ to bring the action back to the original form. We obtain
\begin{eqnarray}
&& Z_\psi = e^{(2+z_1)\ell},\\
&& Z_{v_z} = e^{(z-z_1)\ell}\left[ 1 + \frac{\sqrt{\eta}e^2 \ell}{4\pi^2v_z}\right],\\
&& Z_{m^{-1}} = e^{(z-2)\ell} \left[ 1 + \frac{c \sqrt{\eta}e^2}{4\pi^2v_z}\ell\right],\\
&& Z_\phi = e^{(z+1) \ell} \left[ 1 + \frac{\sqrt{\eta} e^2}{3\pi^2 v_z} \ell\right]^{\frac{1}{2}} \left[ 1 + \frac{m^2 v_z e^2}{12 \pi^2 \sqrt{\eta}\Lambda^2}\ell\right]^{\frac{1}{2}},\\
&& Z_\eta = e^{(1-z_1)\ell} \left[ 1 + \frac{m^2 v_z e^2}{12\pi^2\sqrt{\eta}\Lambda^2}\ell\right]\left[1+\frac{\sqrt{\eta}e^2}{3\pi^2 v_z}\ell\right]^{-1},\\
&& Z_{e^2} = e^{(z-1)\ell} \left[ 1 + \frac{\sqrt{\eta}e^2}{3\pi^2v_z}\ell\right]^{-\frac{1}{2}}\left[ 1+ \frac{m^2 v_z e^2}{12\pi^2 \sqrt{\eta}\Lambda^2}\ell\right]^{-\frac{1}{2}}. \label{Eq:Ze}
\end{eqnarray}
The RG equations are
\begin{eqnarray}
&& \frac{d \ln v_z}{d\ell} = z - z_1 +\frac{\sqrt{\eta}e^2\ell}{4\pi^2 v_z},\\
&& \frac{d \ln m^{-1}}{d\ell} = \left( z -2 + \frac{c \sqrt{\eta}e^2}{4\pi^2 v_z}\right),\\
&& \frac{d \ln \eta}{d\ell} = 2 (1-z_1) + \frac{m^2 v_z e^2}{12\pi^2 \sqrt{\eta} \Lambda^2} - \frac{\sqrt{\eta} e^2}{3\pi^2 v_z},\\
&& \frac{d \ln e^2}{d\ell} =  z-1 -\frac{\sqrt{\eta}e^2}{6\pi^2 v_z} - \frac{m^2 v_z e^2}{24\pi^2 \sqrt{\eta} \Lambda^2}.
\end{eqnarray}
Introducing the dimensionless parameters, 
\begin{equation}
\begin{array}{lr}
\alpha \equiv \frac{me^2}{12\pi^2 \Lambda}, &\lambda \equiv \frac{m^2 v_z e^2}{48\pi^2 \sqrt{\eta}\Lambda^2},
\end{array}
\end{equation}
we obtain the RG equations,
\begin{eqnarray}
&& \frac{d \ln v_z}{d\ell} = z - z_1 +\frac{3}{4}\frac{\alpha^2}{\lambda},\\
&& \frac{d \ln m^{-1}}{d\ell} = z -2 + \frac{3c}{4}\frac{\alpha^2}{\lambda},\\
&& \frac{d \ln \alpha}{d\ell} = z-1 - 2\lambda - \frac{1}{2}\frac{\alpha^2}{\lambda},\\
&& \frac{d \ln \lambda}{d\ell} =  z + z_1 -2 - 4\lambda.
\end{eqnarray}
If we hold $v_z$ and $m^{-1}$ fixed, we get
\begin{eqnarray}
&& z= 2 - \frac{3c}{4} \frac{\alpha^2}{\lambda},\\
&& z_1 = z + \frac{3}{4}\frac{\alpha^2}{\lambda} = 2 + \frac{3(1-c)}{4}\frac{\alpha^2}{\lambda}.
\end{eqnarray}
The RG equations for double-Weyl semimetals in the presence of long-range Coulomb interaction are
\begin{eqnarray}
&& \frac{d\alpha}{d\ell} = \alpha \left[ 1 - 2\lambda  - \frac{3c+2}{4}\frac{\alpha^2}{\lambda}\right],\\
&& \frac{d \lambda}{d\ell} = 2\lambda  \left[ 1 - 2\lambda  - \frac{6c-3}{8}\frac{\alpha^2}{\lambda}\right].
\end{eqnarray}
We can see that fixed points are located at $( \alpha,~\lambda) = (0,~0)$, and $(0,~1/2).$ Linearizing around these two fixed points, we find that the fixed point $(0,~0)$ is the {\it unstable} Gaussian fixed point, and $(0,~1/2)$ is the stable fixed point controlled parameter defined as the ratio of long-range Coulomb interaction and the anisotropic parameter. The RG flow diagram is shown in Fig.~2 in the main texts.

\section{Random Phase Approximation analysis of screened Coulomb interaction in double-Weyl semimetals}\label{APP:RPA}
We use RPA analysis to examine the screened Coulomb interaction in double-Weyl semimetals. We will focus on polarization function $\Pi(\vec{k},\omega)$ illustrated in Fig.~2(b) in the main texts and perform the integral without restricting integrating range. For clarity, we relabel the frequency and the momenta $(\omega, k_x,k_y,k_z) \rightarrow (k_0, k_1,k_2,k_3)$. The polarization function after proper scaling of the variable is
\begin{eqnarray}
\nonumber && -\frac{v_z}{2me^2} \Pi(k_0, \sqrt{m} \vec{k}_\perp, \frac{k_3}{v_z}) \\
\nonumber && =  \int_q \frac{q_0(q_0+k_0) -  q_3 (q_3 + k_3) - d_j(\vec{q}-\frac{\vec{k}}{2}) d_j(\vec{q} + \frac{\vec{k}}{2})}{\left[ q_0^2 +E_q^2 \right] \left[ \left( q_0 + k_0 \right)^2 + E_{q+k}^2 \right]} \\
\nonumber && =  \int_{q,x} \frac{q_0(q_0 + k_0) - q_3 (q_3 + k_3) -d_j(\vec{q}-\frac{\vec{k}}{2}) d_j(\vec{q} + \frac{\vec{k}}{2})}{\left[ \left( \mathbf{q}+x \mathbf{k} \right)^2 + x(1-x) \mathbf{k}^2 + x (q_\perp^+)^4 + (1-x) (q_\perp^-)^4\right]^2},\\
\end{eqnarray}
where $\int_{q,x} = \int_q \int_0^1 dx$ and we introduced the Feynman parameter $x$, which leads to two vectors $\mathbf{q} =(q_0, q_3)$, $\mathbf{k} = (k_0, k_3)$ and $(q_\perp^{\pm})^2\equiv (q_1 \pm k_1/2)^2 + (q_2\pm k_2/2)^2$ and the repeated subscript indices $j$ means summation over $j=1,2$. We also introduce the rescaled dispersion $E_k^2 =  k_\perp^4 + k_z^2$. We then introduce $\bar{\mathbf{q}} = \mathbf{q} + x \mathbf{k}$ so that $\mathbf{q} ( \mathbf{q} + \mathbf{k}) = (\bar{\mathbf{q}} - x \mathbf{k}) \left[ \bar{\mathbf{q}} + (1-x)\mathbf{k}\right]$ and perform the integration of $q_0$ and $q_3$. We obtain the static polarization function as
\begin{eqnarray}
\nonumber && \Pi(0, \sqrt{m} \vec{k}_\perp, \frac{k_3}{v_z} ) \\
\nonumber &&= - \frac{me^2}{8\pi^3 v_z} \int_0^1 dx \int_{q_\perp}  \frac{-d_j (q_\perp^+) d_j(q_\perp^-) + x(1-x) k_3^2}{x(1-x) k_3^2 + x (q_\perp^+)^4 + (1-x) (q_\perp^-)^4}.\\
\end{eqnarray}
Now we can examine the leading terms in $k_3$ and $k_\perp$. First if we set $k_\perp=0$, the result after regularization is
\begin{eqnarray}
\Pi(0, 0, \frac{k_3}{v_z}) = -\frac{\pi m e^2}{64 v_z} |k_3| \propto |k_3|,
\end{eqnarray}
which is \textit{linear} in $k_3$. If we set $k_3 = 0$, the integral can not be performed analytically. But we can factorize out the $k_\perp$ dependence to see how the result scales with $k_\perp$. We find that the result is
\begin{eqnarray}
 \Pi(0,\sqrt{m} \vec{k}_\perp,0) =- \frac{me^2}{4\pi^3 v_z} k^2_{\perp}  \int_{x, y} f(x,y) \propto k_\perp^2, \label{Eq:pfperp}
\end{eqnarray}
where 
\begin{eqnarray}
\nonumber f(x,y)=&&\frac{\frac{1}{16} + x^4 + y^4 - \frac{3}{2} y^2 + x^2(2 y^2 -\frac{1}{2})}{4x^3 + 4 x y^2 + x} \times \\
&& \hspace{2.5cm}\times \ln\left|\frac{(2x-1)^2 + 4 y^2}{(2x+1)^2 + 4y^2}\right|.
\end{eqnarray}
We can see from Eq.~(\ref{Eq:pfperp}) that the leading term in $k_\perp$ is still quadratic. We can conclude that the leading terms in RPA analysis is
\begin{eqnarray}
\Pi(0,\sqrt{m}\vec{k}_\perp,\frac{k_z}{v_z}) \sim k_\perp^2 + |k_z|,
\end{eqnarray}
which is consistent with the RG analysis. 

\section{Large-$N_f$ analysis}\label{APP:Large_N}
In this appendix, we will illustrate how the infrared divergency near the stable fixed point arises via a simplified large $N_f$ analysis. The strong-coupling analysis starts with the action at the stable fixed point,
\begin{eqnarray}
S_s = \int \psi^\dagger\left(H_0 -i\phi\right)\psi + N_f \int \left(q_\perp^2 + \left| q_3 \right|\right) \left|\phi_{q,\omega}\right|^2,
\end{eqnarray}
where $N_f$ different copies of fermions are introduced and we suppress dimensionful numbers and the boson propagator is from the RPA calculation. Here instead of evaluating the $N_f\rightarrow \infty$ limit exactly, we use the RPA results that capture the correct momentum dependence in each direction. Since we are only interested in how the infrared divergence appears, the use of the simplified RPA result can be justified. It is important that the electric charge does not appear in the action since it is always possible to absorb the constant into the boson field by redefining the field. 

Given the approximate action, we can evaluate $1/N_f$ correction. The electron self-energy with the momentum cutoff $\Lambda$ in the quadratic direction is
\begin{eqnarray}
\Sigma_f(k,\omega) = \frac{1}{N_f} \int_q \frac{d_a(k+q)\sigma^a}{E_{k+q}}\frac{1}{q_\perp^2 + |q_3|},
\end{eqnarray}
where $d_a \sigma^a \sim (k_1^2 - k_2^2) \sigma^1 + 2 k_1 k_2 \sigma^2 + k_3 \sigma^3$ and the dispersion $E(k) = \sqrt{\sum_a (d_a)^2}$. The correction can be read off by considering the $k\rightarrow 0$ limit as
\begin{eqnarray}
\frac{\partial \Sigma_f(k+q)}{\partial d_a}\bigg{|}_{k\rightarrow 0} \sim \frac{1}{N_f} \int_{\mu^2}^{\Lambda_\perp^2} \frac{dq_\perp^2}{q_\perp^2} \sim \ln \frac{\Lambda_\perp^2}{\mu^2}.
\end{eqnarray}
We can see there is a logarithmic divergence at the infrared limit $(\mu \rightarrow 0)$. Hence, the anisotropic screening induces the logarithmic correction similar to conventional graphene physics.

\section{Dynamic conductivity at noninteracting limit}\label{APP:dynamic_conductivity}
Within linear response theory, we first start from the Matsubara formalism to calculate the current-current correlation function
\begin{eqnarray}
\nonumber \Pi_{\mu \mu} (i\omega_n) = -e^2 \int_{0}^{\beta} d\tau e^{i\omega_n \tau} \left\la T_\tau \left[ J_\mu (\tau) J_\mu (0)\right] \right\ra,\\
\end{eqnarray}
and  perform analytic continuation to the real frequency as
\begin{eqnarray}
\Pi_{\mu \mu} (\omega) = \Pi_{\alpha\beta}(i\omega_n \rightarrow \omega+ i\eta^+).
\end{eqnarray}
In the end, we can extract the dynamic conductivity by extracting the imaginary part
\begin{eqnarray}
\sigma_{\mu \mu}(\omega) = - \frac{ Im \left[ \Pi_{\mu\mu}(\omega)\right]}{\omega}.
\end{eqnarray}
Before the indulging in the calculations, we first note that due to the rotation symmetry, $\Pi_{xx} (i \omega_n) = \Pi_{yy}(i \omega_n) \not= \Pi_{zz}(i \omega_n)$. We will discuss separately $\Pi_{xx}(i\omega_n)=\Pi_{yy}(i\omega_n)$ and $\Pi_{zz}(i\omega_n)$, which lead to the conductivity $\sigma_{xx}(\omega)=\sigma_{yy} (\omega) \not =\sigma_{zz}(\omega)$.

The current components are
\begin{eqnarray}
&& J_x = \int_q \psi^\dagger_q \left( \frac{2q_x}{m} \sigma^x + \frac{2q_y}{m}\sigma^y\right)\psi_q \equiv \int_q \psi_q^\dagger\ \mathcal{J}_x\ \psi_q,\\
&& J_y = \int_q \psi^\dagger_q \left( -\frac{2q_y}{m} \sigma^x + \frac{2q_x}{m}\sigma^y\right)\psi_q \equiv \int_q \psi^\dagger_q\ \mathcal{J}_y\ \psi_q,~~\\
&& J_z = \int_q \psi^\dagger_q v_z \sigma^z \psi_q \equiv \int_q \psi^\dagger_q\ \mathcal{J}_z \ \psi_q.
\end{eqnarray}
and  the diagonal current-current correlation function in the Matsubara domain can be expressed as
\begin{eqnarray}
\nonumber \Pi_{\mu\mu}(i\omega_n) = \frac{e^2}{\beta} \sum_{m} \int_q Tr \left[\mathcal{J}_\mu \mathcal{G}(\vec{q},ip_m) \mathcal{J}_\mu \mathcal{G}(\vec{q}, ip_m + i \omega_n)\right],\\
\end{eqnarray}
where $\mathcal{G}(\vec{q}, i\omega_n)$ is the noninteracting fermion green's function in the Matsubara domain as
\begin{eqnarray}
\nonumber \mathcal{G}({\bf k}, i \omega_n) = \frac{1}{i\omega_n + \mu - \mathcal{H}_0},\\
\end{eqnarray}
where the $\mathcal{H}_0$ is the Hamiltonian density of the system. After straightforward derivation, below we list the main results.\\
\textbf{(A) Drude weight at zero frequency,} $\sigma_{\mu \mu}(\omega)\bigg{|}_{\omega \rightarrow 0}$:
\begin{eqnarray}
\nonumber  && \sigma_{xx/yy}(\omega)\bigg{|}_{\omega \rightarrow 0}=\\
\nonumber &&  = -\frac{e^2}{3\pi m v_z}(2mT) \left[ Li_2 (- e^{\mu/T} )+ Li_2(-e^{-\mu/T})\right] \delta \left( \frac{\omega}{T} \right),\\
\\
\nonumber && \sigma_{zz} (\omega)\bigg{|}_{\omega \rightarrow 0} = \\
&& = - \frac{mv_ze^2}{16} \left[ Li_1(-e^{\mu/T}) + Li_1(-e^{-\mu/T})\right] \delta\left( \frac{\mu}{T} \right).
\end{eqnarray}
The Drude weights show different behaviors at different limits.\\
\textbf{(1) $\mu/T \ll 1$}:
\begin{eqnarray}
&& \sigma_{xx/yy}(\omega )\bigg{|}_{\omega\rightarrow 0} \rightarrow \frac{\zeta(2)e^2}{3\pi m v_z} (2mT) \delta \left( \frac{\omega}{T}\right);~~\\
&& \sigma_{zz}(\omega )\bigg{|}_{\omega\rightarrow 0} \rightarrow \frac{ mv_z e^2}{8} \ln(2) \delta \left( \frac{\omega}{T}\right).
\end{eqnarray}\\

\textbf{(2) $ \mu/T \gg 1$}:
\begin{eqnarray}
&& \sigma_{xx/yy}(\omega)\bigg{|}_{\omega \rightarrow 0} \rightarrow \frac{e^2}{6\pi m v_z} (2mT) \left( \frac{\mu}{T}\right)^2 \delta\left( \frac{\omega}{T}\right);\\
&& \sigma_{zz}(\omega)\bigg{|}_{\omega \rightarrow 0} \rightarrow \frac{mv_z e^2}{16} \left(\frac{\mu}{T}\right) \delta \left( \frac{\omega}{T}\right).
\end{eqnarray}

\textbf{(B) Dynamic conductivity at finite frequency,} $\sigma_{\mu\mu}(\omega \not= 0)$:
\begin{eqnarray}
\nonumber && \sigma_{xx/yy}(\omega)\bigg{|}_{\omega>0} \\
\nonumber &&=\frac{e^2}{24\pi m v_z} (m\omega)\left[ \tanh \left( \frac{\omega - 2 \mu}{4T} \right) + \tanh\left( \frac{\omega + 2 \mu}{4T}\right) \right];~~\\ \\
\nonumber && \sigma_{zz}(\omega)\bigg{|}_{\omega>0} \\
&& = \frac{mv_z e^2}{128}\left[\tanh\left( \frac{\omega - 2 \mu}{4T} \right) + \tanh\left( \frac{\omega + 2 \mu}{4T}\right) \right].
\end{eqnarray}
Combining both the zero frequency and finite frequency parts, the dynamic conductivity can be expressed as $\sigma_{xx}(\omega, T) =\sigma_{yy}(\omega,T)= \frac{e^2}{3\pi m v_z} (2mT) \Phi_\perp(\omega/T, \mu/T)$, and $\sigma_{zz}(\omega, T) = \frac{mv_z e^2}{16} \Phi_z(\omega/T,\mu/T)$ with the scaling function
\begin{eqnarray}
\nonumber  \Phi_\perp(a,b) &=& -\left[ Li_2 (-e^{b})
+ Li_2 ( - e^{-b}) \right]\delta(a) +\\
&& + \frac{1}{16} a \left[ \tanh (\frac{a}{4} + \frac{b}{2}) + \tanh (\frac{a}{4} - \frac{b}{2})\right];~~~~\\
\nonumber \Phi_z(a,b) &=& -\left[ Li_1 (-e^{b})
+ Li_1 ( - e^{-b}) \right]\delta(a)  + \\
&&+ \frac{1}{8} a \left[ \tanh (\frac{a}{4} + \frac{b}{2}) + \tanh (\frac{a}{4} - \frac{b}{2})\right].
\end{eqnarray}

\section{Diamagnetic susceptibility at noninteracting limit}\label{APP:diamagnetic_sus}
In the noninteracting limit, the Hamiltonian is 
\begin{eqnarray}
H_{DWS} = \frac{q_x^2-q_y^2}{m}\sigma_x + \frac{2q_x q_y}{m}\sigma_y + v_z q_z \sigma_z.
\end{eqnarray}
In order to calculate the diamagnetic susceptibility, we will use the Fukuyama formula as
\begin{eqnarray}
\chi_D = e^2 \frac{1}{\beta} \sum_n \int_q Tr\left[ \mathcal{G}_0\gamma_a \mathcal{G}_0\gamma_b \mathcal{G}_0\gamma_a \mathcal{G}_0\gamma_b\right],
\end{eqnarray}
where $\mathcal{G}_0$ is the fermion Green's function in the Matsubara domain, the $n$ summation represents the Matsubara frequency sum, and $\gamma_a \equiv \partial H/\partial q_a$, with $a$ being the direction axis that perpendicular to the direction of the magnetic field. Before we go into the calculations, we first note that due to the anisotropy it is expected that the diamagnetic susceptibilities for the cases with $\vec{B} = B \hat{x}$ and $\vec{B}= B\hat{x}$ are the same. However, the diamagnetic susceptibility for the case with $\vec{B} = B \hat{z}$ should be different to the two former cases. Let us discuss each case separately below to see the temperature dependence of the diamagnetic susceptibilities in different cases. Below, we set the chemical potential to be zero.

First, we choose $\vec{B} = B \hat{x}$ and the result should be the same to the case of $\vec{B} = B \hat{y}$. Now, we have $\gamma_y = -(2q_y/m) \sigma_x + (2 q_x/m \sigma_y)$ and $\gamma_z = v_z \sigma_z$. The Fukuyama formula gives
\begin{eqnarray}
\chi_D^{\perp} &=& e^2 \frac{1}{\beta} \sum_n \int_q Tr \left[ \mathcal{G}_0\gamma_y \mathcal{G}_0 \gamma_z \mathcal{G}_0 \gamma_y \mathcal{G}_0 \gamma_z\right] \\
\nonumber & = & \frac{4e^2v_z^2}{m^2}\frac{1}{\beta} \sum_n \int_q Tr\bigg{[} q_x^2 \mathcal{G}_0\sigma_y \mathcal{G}_0 \sigma_z \mathcal{G}_0 \sigma_y \mathcal{G}_0 \sigma_z +\\
\nonumber &&\hspace{2.8cm}+ q_y^2 \mathcal{G}_0\sigma_x \mathcal{G}_0 \sigma_z \mathcal{G}_0 \sigma_x \mathcal{G}_0 \sigma_z\bigg{]}\\
& = &\chi_D^{\perp,(I)} + \chi_D^{\perp,(II)}. 
\end{eqnarray}
After expansion and exchanging $q_x$ and $q_y$ for $\chi_D^{\perp,(II)}$, we find $\chi_D^{\perp, (II)} = \chi_D^{\perp,(I)}$. After performing the Matsubara frequency summation, we get
\begin{widetext}
\begin{eqnarray}
\nonumber  \chi^\perp_D =- \frac{16e^2 v_z^2}{m^2} \int_q &\bigg{\{}&q_x^2\bigg{[} \frac{ \tanh (\frac{d}{2T}) }{4d^3}+ \frac{sech^2(\frac{d}{2T})}{8d^2T} \bigg{]} +\\
&+& 8 q_x^2 d_2^2 d_3^2 \bigg{[} \frac{sech^2(\frac{d}{2T})\tanh^2(\frac{d}{2T}) }{192 d^4 T^3}+ \frac{sech^2(\frac{d}{2T})\tanh(\frac{d}{2T})}{32d^5T^2} +\frac{5 sech^2(\frac{d}{2T})}{64 d^6 T}  - \frac{5 \tanh(\frac{d}{2T})}{32d^7}\bigg{]} \bigg{\}}.
\end{eqnarray}
\end{widetext}
The momentum integral is complicated, but since we are only interested in the temperature dependence, we can factorize out the temperature dependence by rescaling
\begin{eqnarray}
q_x = (mT)^{\frac{1}{2}} x,~q_y = (mT)^{\frac{1}{2}} y,~q_z = T v_z^{-1} z.
\end{eqnarray}
After the rescaling and straightforward algebra, we get the diamagnetic susceptibility for $\vec{B} = B \hat{x}$,
\begin{widetext}
\begin{eqnarray}
\nonumber && \chi^\perp_D =- e^2 v_z \int_{-\infty}^{\infty} \frac{dx dy dz}{(2\pi)^3}\bigg{\{}2x^2 \bigg{[} \frac{2 \tanh(\frac{\bar{d}}{2})}{\bar{d}^3} + \frac{sech^2(\frac{\bar{d}}{2})}{\bar{d}^2}  \bigg{]} + \\
\nonumber && \hspace{3cm} + \frac{2}{3}x^2 \bar{d}_2^2 \bar{d}_3^2 \bigg[ \frac{sech^2(\frac{\bar{d}}{2})\tanh^2(\frac{\bar{d}}{2})}{\bar{d}^4} + \frac{6 sech^2(\frac{\bar{d}}{2}) \tanh(\frac{\bar{d}}{2})}{\bar{d}^4}+ \frac{15 sech^2(\frac{\bar{d}}{2})}{\bar{d}^6} - \frac{30 \tanh(\frac{\bar{d}}{2})}{\bar{d}^7} \bigg{]}\bigg{\}} \\
&& \sim -e^2v_z,
\end{eqnarray}
\end{widetext}
where we introduce $\bar{d} \equiv \left(\bar{d}_1^2 +\bar{d}_2^2 + \bar{d}_3^2\right)^{1/2}$, and $ \bar{d}_1 = x^2-y^2$, $\bar{d}_2 = 2xy$, and $\bar{d}_3 = z$. The result above should be the same for the case with $\vec{B} = B \hat{y}$. Now let us check the temperature dependence for the diamagnetic susceptibility in the presence of $\vec{B} = B \hat{z}$. In this case, we need $\gamma_x = (2q_x/m) \sigma_x + (2q_y/m)\sigma_y$, and $\gamma_y = -(2q_y/m)\sigma_x + (2q_x/m) \sigma_y$.  The Fukuyama formula gives
\begin{eqnarray}
\chi_D^z = e^2\frac{1}{\beta} \sum_n \int_q Tr \bigg{[} \mathcal{G}_0 \gamma_x \mathcal{G}_0 \gamma_y \mathcal{G}_0 \gamma_x \mathcal{G}_0 \gamma_y \bigg{]}.
\end{eqnarray}
After expansion and performing Matsubara frequency summation, we get
\begin{widetext}
\begin{eqnarray}
\nonumber  \chi_D^z = && -\frac{32e^2}{m^4}\int_q \bigg{[} \frac{q_\perp^4\tanh (\frac{d}{2T})}{4d^3} + \frac{q_\perp^4 sech^2(\frac{d}{2T})}{8d^2 T}\bigg{]} \\
\nonumber && - \frac{256e^2}{m^8} \int_q q_x^2q_y^2 \bigg{[} q_\perp^8 + 4(q_x^2-q_y^2)^4\bigg{]}
 \bigg{[} \frac{sech^2(\frac{d}{2T})\tanh^2(\frac{d}{2T})}{192d^4 T^3} + \frac{sech^2(\frac{d}{2T})\tanh(\frac{d}{2T})}{32d^5 T^2} + \\
 && \hspace{9cm} + \frac{5 sech^2(\frac{d}{2T})}{64 d^6 T} - \frac{5\tanh(\frac{d}{2T})}{32d^7}\bigg{]}.
\end{eqnarray}
\end{widetext}
We can again factorize out the temperature dependence. We find 
\begin{widetext}
\begin{eqnarray}
\nonumber \chi_D^z &= & - \frac{4e^2 T}{m v_z} \int_{-\infty}^{\infty} \frac{dx dy dz}{(2\pi)^3} \bigg{[} \frac{2 \tanh(\frac{\bar{d}}{2})}{\bar{d}^3} + \frac{sech^2(\frac{\bar{d}}{2})}{\bar{d}^2}\bigg{]} - \\
\nonumber && - \frac{4e^2 T}{3m v_z} \int_{-\infty}^{\infty} \frac{dx dy dz}{(2\pi)^3} x^2 y^2 \bigg{[} r_\perp^8 + 4 (x^2 -y^2)^4 \bigg{]} \bigg{[} \frac{sech^2(\frac{\bar{d}}{2})}{\bar{d}^4}\tanh^2(\frac{\bar{d}}{2})+ \frac{6 sech^2(\frac{\bar{d}}{2})\tanh(\frac{\bar{d}}{2})}{\bar{d}^5} + \\
\nonumber&& \hspace{9cm} + \frac{15 sech^2(\frac{\bar{d}}{2})}{\bar{d}^6} -\frac{30 \tanh(\frac{\bar{d}}{2})}{\bar{d}^7}\bigg{]} \\
&& \sim - \frac{e^2 T}{mv_z}.
\end{eqnarray} 
\end{widetext}
Hence, in the presence of $\vec{B} = B\hat{z}$ the diamagnetic susceptibility is actually linearly proportional to the temperature. Combining the results of the cases of $\vec{B} = B \hat{r}_\perp$ and $\vec{B} = B \hat{z}$, we expect that the diamagnetic susceptibility in the presence of magnetic field in arbitrary direction $\vec{B} = B \hat{r}$ should show the temperature dependence as
\begin{eqnarray}
\chi_D \sim \left(\sin^2 \theta~\chi_0 + \cos^2 \theta~T\right),
\end{eqnarray}
where $\chi_0$ is a constant independent of $T$ and we introduce the periodic function with $\theta$ being the angle between the magnetic field and the $\vec{z}-$axis, i.e. $\vec{B}\cdot \vec{z} = B \cos\theta$. The square of the periodic functions roughly gives the correct angular dependence with $\chi_D (\theta) = \chi_D(\theta+\pi)$. Therefore, at low temperature limit $T \rightarrow 0$ we expect that the constant diamagnetic susceptibility dominates.


\bibliography{biblio4MWeyl}

\end{document}